\documentclass[12pt]{article}
\usepackage{graphicx}
\usepackage{amssymb}
\usepackage{epstopdf}
\usepackage{multirow}
\newcommand\mathC{\mkern1mu\raise2.2pt\hbox{$\scriptscriptstyle|$}
        {\mkern-7mu\rm C}}              

\newcommand{\be}{\begin{equation}}
\newcommand{\ee}{\end{equation}}

\let\ssection=\section
\renewcommand{\section}{\setcounter{equation}{0}\ssection}

\newcommand{\raw}{\rightarrow}

\def\Dslash{\setbox0=\hbox{$D$}D\hskip-\wd0\hbox to\wd0{\hss\sl/\/\hss}}

\begin{document}
\begin{center}
{\large\bf On Under-determination in Cosmology}
\end{center}

\begin{center}
Jeremy Butterfield\\
Trinity College, Cambridge CB2 1TQ: jb56@cam.ac.uk
\end{center}

\begin{center}
Published in {\em Studies in History and Philosophy of Modern Physics} {\bf 46} (2014), pp. 57-69
\end{center}

\begin{center}
\end{center}

\begin{center}  5 July 2013 \end{center}

\begin{abstract}
I discuss how modern cosmology illustrates under-determination of theoretical hypotheses by 
data, in ways that are different from most philosophical discussions. I emphasize 
cosmology's concern with what data could in principle be collected by a single observer 
(Section \ref{obsind}); and I give a broadly sceptical discussion of cosmology's appeal to 
the cosmological principle as a way of breaking the under-determination (Section \ref{CP}).

I confine most of the discussion to the history of the observable universe from about one 
second after the Big Bang, as described by the mainstream cosmological model: in effect, 
what cosmologists in the early 1970s dubbed the `standard model', as elaborated since then. 
But in the closing Section \ref{env}, I broach some questions about times earlier than one 
second.

\end{abstract}

\newpage
\tableofcontents

\newpage

\section{Introduction}\label{intro}

\subsection{Prospectus: two differences}\label{prosp}
The aim of this paper is to survey how modern cosmology illustrates the under-determination 
of theoretical hypotheses by data in ways that are different from most philosophical 
discussions. I will concentrate on two main differences, which I can introduce as follows.

The usual philosophical discussion of under-determination idealizes by assuming that all 
possible observations are `given', even though of course they are spread throughout 
spacetime, and so never collected by a single scientific community. Then the discussion 
addresses whether a theory of tiny unobservable objects like electrons is under-determined; 
and relatedly, how good is our warrant for our accepted physical theories. 

On the other hand, cosmology takes due cognizance of the difficulties both of observing 
parts of the universe that are very distant in time and space, and of collecting 
observations together at a single place and time. But (as one would expect): cosmology does 
not address the general philosophical debate about our warrant for our accepted theories, in 
particular our current theories of tiny unobservable objects. On the contrary, cosmology for 
the most part treats  our current accepted physical theories as `given'. For it is, like 
geology, a historical science: and as such, it aims to provide, not a general theory, but as 
detailed as possible a history of its topic---the universe.\footnote{My usage here of 
`historical science' is of course not meant to exclude prediction, the most striking current 
example being no doubt the universe's accelerating expansion. My usage just indicates the 
contrast between a general theory admitting many models (solutions, `possible worlds') and a 
detailed description of a single model. Agreed, the usage of `theory' and `model' varies: 
for clarification, cf. below, especially (2) in Section \ref{scene}.}

Here I say `for the most part' because, although cosmology has in the last several decades 
gone from success to success in describing, in ever more detail, the history of
the  universe, using our accepted theories, especially general relativity and quantum 
theory---nevertheless, this endeavour of course leads to unknown physics, especially at the 
very high energies in the very early universe. I myself would place the limit of justified 
confidence in applying our accepted theories at about one second after the Big Bang (a 
temperature of about $10^{10}$ K); and  I will confine my discussion to that and later times 
(lower energies)---except for my last Section which poses, but does not pursue, some 
questions about earlier times.\footnote{Placing the limit at one second is cautious. 
Nowadays, it is common to take the boundary between known and speculative physics to be  at 
about $10^{-11}$ seconds, before which the energies are too high for us to be confident that 
the  standard model of particle
physics applies. This time corresponds to the electro-weak phase transition. And even twenty 
years ago, a standard monograph was confident of its account from $10^{-2}$ seconds onwards 
(Kolb and Turner, 1990, p. 109). Cf. also  Rees' remark  that he is 99\% confident of the 
account from one second onwards (2003, pp. 24, 31; 1997, p. 65, 174). For a philosophical 
entry into these details, cf. Zinkernagel (2002), Rugh and Zinkernagel (2009), Smeenk (2012, 
Section 6).
My last Section will broach the idea of inflation, which is meant to have started much 
(logarithmically)
earlier than $10^{-11}$ seconds, viz. at about $10^{-35}$ seconds: so I think caution about 
inflation is all the more justified. On the other hand, I also think that both this 
confrontation with unknown physics, and the under-determination which this paper is 
concerned with, are entirely compatible with scientific realism: I argue for this elsewhere 
(2012, Section 2).}

Thus I see two main differences. The first is that cosmology takes in to account the task of 
collecting at a single place and time all the possible data ---which most philosophical 
discussions do not. When we add to this the supreme success of modern cosmology in 
describing the history of the universe from about one second after the Big Bang, we are led 
to the second difference. Namely: in cosmology, general theories are {\em not}, for the most 
part, the items that are under-determined. Rather what is under-determined is which  {\em 
model} of our well-established theories is the correct one; i.e. which history of the 
universe as described by general relativity and quantum theory should we endorse. (Agreed, 
this difference is less crisp than the first, not least because the theory-model contrast is 
vague; cf. Section \ref{scene}.)

The first of these differences is entrenched in a common jargon within cosmology---which I 
will adopt. Thus cosmologists often use {\em observable universe} to mean (roughly speaking) 
`the past light-cone of Earth-now, and all  physical events within it, even microscopic (and 
so humanly unobservable) ones'; as against, say, `all macroscopic or humanly observable 
events, anywhere in spacetime'. Combining this jargon with the second difference, we 
conclude: in cosmology, the issue of under-determination is often, not about our warrant for 
our general theories, but about  whether the data about the observable universe determine a 
single  model of these theories as being correct.

I will proceed in three main stages. In Section \ref{obsind}, I will discuss recent results 
of Manchak (2009, 2011): their gist will be that according to general relativity, the 
correct model is endemically under-determined by data, even ideal data, about the observable 
universe (i.e. past light-cone of an observer). This predicament prompts the question 
whether we can justify some principle that cuts down the dismaying variety of models. The 
cosmological principle is probably the strongest and best known candidate for such a 
principle. So in Section \ref{CP}, I survey how we might go about justifying it, by general 
arguments and-or by empirical evidence. My verdict will be mostly negative: although the 
cosmological principle holds up well as an idealized description of the observable universe, 
i.e. our past light-cone, it is very hard to justify for the universe as a whole. I will 
end, in Section \ref{env}, by posing (but not pursuing) some questions that arise, including 
questions about  times earlier than one second after the Big Bang (and correspondingly 
higher energies). But first: I should set the scene, by briefly discussing some other 
aspects of under-determination  in cosmology.

\subsection{Setting the scene}\label{scene}
In cosmology, under-determination is not always understood in the way I have sketched. Hence 
I said `often', rather than `always', when I reported how cosmologists use `observable 
universe', and when I said that under-determination is about whether data about the 
observable universe determine a single  model. 

As I see matters, there are two other main construals of `observable universe', and-or other 
relevant terms like `under-determination'. It will help to state them explicitly: not least 
because they are both close to common philosophical concerns. The first brings us closer to 
philosophers' usual theory-observation distinction. The second registers the variety of uses 
of `theory' and `model'.

\indent (1): {\em `Observable' in the light of decoupling}: Thus cosmologists also use 
`observable universe' more narrowly, reflecting their concern with what can be ascertained 
by observation. The last several decades have of course seen great advances in cosmological 
observation: both in the types of radiation (parts of the electromagnetic spectrum) and of 
matter observed, and in observations' precision and depth; (`depth' in time as well as 
space: cf. e.g. Longair (2003, Chapter 18; 2006 Chapters 7, 11, 13). The most famous advance 
has been the cosmic microwave background radiation (predicted in 1950 and discovered in 
1965) which dates from the decoupling time, about 400,000 years after the Big Bang: and 
which is nowadays our single most important line of evidence for the Big Bang, and about the 
early universe---witness the data about it from the COBE, WMAP and Planck projects. But we 
cannot now directly observe by any  electromagnetic means any earlier events, since until 
the decoupling time
the universe was opaque to radiation. So until and unless we observe e.g. gravitational 
radiation from such earlier times, cosmologists often take the phrase `observable universe' 
to mean: `the events in the past light-cone of Earth-now, and to the future of the 
decoupling time, that our instruments can detect'. With this sort of usage, the issue of 
under-determination of theory by observation in cosmology becomes closer to the familiar 
philosophical issue of the fallibility of inference from the observed to the unobserved.

\indent (2): {\em `Theory' vs. `model'}: I said that cosmology treats our general theories, 
such as general relativity and quantum theory, as `given'; and using them, it has provided a 
very successful, ever more detailed, history of the universe. And I qualified this, adding 
`for the most part', only by registering our ignorance of the very early universe. (This 
qualification is similar to the shift from my first, adopted, sense of `observable 
universe', to that in (1) above.)

Broadly speaking, that is true enough. Consider how the overall thermal history of the 
observable  universe has been well established since about 1970. We know that about 13 
billion years ago, there was an extremely hot and dense `fireball' in which light nuclei 
(like hydrogen and helium) were synthesized, according to well-understood nuclear physics; 
it cooled and expanded, with the details being accurately described by quantum theory, 
thermodynamics and general relativity; later, gravitational clustering led to galaxies and 
stars, in which still more elements were synthesized.\footnote{{\label{thermal}}{Thus 
authoritative
textbook descriptions of this thermal history, written over the last forty years, largely 
agree with each other. Cf. for example: Sciama (1971: Chapters 8, 12-14), Weinberg (1972, 
Chapter 15.6, pp. 528-545), Wald (1984, pp. 107-117), Barrow and Tipler (1988,  pp. 367-408, 
Sections 6.1-6.7), Lawrie (1990, pp. 315-326), Longair (2006, 394-399), Weinberg (2008, pp. 
101-113, 149-173; Sections 2.1, 2.2, 3.1, 3.2). For fine popular accounts, cf.  Silk (1989, 
Chapters 6 to 8), Rowan-Robinson (1999, Chapter 5),  Silk (2006: pp. 112-128).

Besides, the thermal history of the observable universe is by no means the only over-arching 
cosmological claim that is now firmly established. The theory of stellar structure and 
evolution is another example, for which one could similarly cite authoritative descriptions 
over several decades largely agreeing with each other.

This consensus is reflected in cosmologists' jargon: what philosophers might call `the Big 
Bang theory' was dubbed by cosmologists
`the standard model', already in the early 1970s (Weinberg (1972, p. 469), Misner et al. 
(1973, p. 763)). At that time, the honorific name reflected its recent confirmation by the 
discovery of
the cosmic background radiation: nowadays, having stood up to many more observational tests, 
it deserves the name all the more.}}

But I should add two other qualifications. Both are independent of issues about the early 
universe; and both give us a warning that the usages of the words `theory' and `model' are 
very variable. (Thanks to C. Beisbart, C.Smeenk, W. Stoeger and H. Zinkernagel for 
emphasizing these to me.)\\
\indent (a): First, in this standard cosmological account (called `the standard model': cf. 
footnote \ref{thermal}), there remain major causal and structural factors that are not 
understood: such as the nature of dark matter and dark energy, and the process of galaxy 
formation. So there are many `versions' of `today's standard model'. In assessing these 
versions, there is much to do, both theoretically and observationally. But as regards the 
philosophical issue of under-determination, the situation seems to me to be `business as 
usual'. (But we must beware the varying jargons: as I have just expressed it, we here face 
`under-determination of versions by presently available data'.)\\
\indent (b) Second: this standard cosmological account, `the standard model', is not 
established beyond doubt: there is respectable dissent. As we will see in more detail in 
Section \ref{CP}, this account incorporates spatial homogeneity (by assuming the 
cosmological principle); but there are rival inhomogeneous models which match the presently 
available data without invoking dark energy (cf. e.g. Sarkar (2007), Nadathur et al (2011)). 
So here again, we have `business as usual': a philosopher would speak of a dispute over---or 
under-determination by presently available data of---theories; and there is good reason to 
hope the dispute will be resolved by observations over the next decade or two.\footnote{For 
more general perspectives on inhomogeneous (and anisotropic) models as tenable rivals to the 
cosmological principle, cf. e.g. MacCallum (1979), Ellis (2007, Section 4.2.2, pp. 
1223-1227; 2011, Section 4, pp. 11-15).

Beware of another variation in jargon. Cosmologists also use `model' for a description of 
(the main features of) a single history, for example in  the name `the concordance 
$\Lambda$CDM  model', where $\Lambda$ refers to the dark energy, and `CDM' refers to `cold 
dark matter'. This model is the best fit of the standard model (an FRW metric as in Section 
\ref{CP11}, and so on) to all observations made so far, and is accordingly very specific 
about the values of cosmological
parameters. It says, in short: 73 per cent dark energy, 27 per cent matter (split between 
baryonic and dark matter, with
about six times more dark matter than baryonic matter), geometrically flat
(with the density parameter $\Omega$ very close to 1) and the Hubble constant $H_0$ about
70 km/sec/Mpc. But to return to (b): I stress that although this concordance model 
represents the current consensus, there are respectable rivals.}

\section{Observationally indistinguishable spacetimes}\label{obsind}
I turn to presenting the endemic under-determination of cosmological models in general 
relativity, by reporting the theorems of Manchak (2009, 2011); (which build on ideas and 
results of Glymour (1977) and Malament (1977)). Roughly, the theorems say that in
almost every spacetime obeying general relativity, no observer, however long they live, 
could accumulate enough observations to exclude their being in another very different 
spacetime.

I begin with some standard notation and jargon. (1): In general relativity, a model is given 
by a spacetime $(M,g)$ consisting of a four-dimensional manifold $M$ of spacetime points, 
equipped with a Lorentzian metric tensor $g$.\\
\indent  (2) A region of one model, $U \subset M$, is {\em isometric} to another region of 
another model, $U' \subset M'$, if there is a diffeomorphism  $d: U \raw U'$  that carries 
$M$'s metric $g$ as restricted to $U$ into $M'$'s metric $g'$ as restricted to $U'$. This 
means full knowledge of all the metrical relations between points in $U$ could not exclude 
one's being instead in $U'$.\\
\indent  (3): Of course, one usually knows much less than everything about metrical 
relations, and more than nothing about spacetime's matter and radiation content: which so 
far, seems culpably unrepresented in our notion, $(M,g)$, of a cosmological model. But the 
theorems and discussion can probably be adapted to include facts about matter and radiation; 
(Malament 1977, pp. 74-76).\footnote{Besides, Einstein's equations for general relativity 
imply that the metric $g$ mathematically determines another tensor $T$, the stress-energy 
tensor, which encodes some main facts about matter and radiation. Hence the widespread 
practice of defining a model by a spacetime $(M,g)$ rather than a triple $(M,g,T)$.}\\
\indent  (4): We formally  define the past light-cone of (an observer at a) spacetime point 
$p \in M$, as follows.  (i): For points $p, q \in M$: $q$ {\em causally precedes} $p$ if 
there is a future-directed curve in $M$ from $q$ to $p$ whose tangent vector at every point 
is timelike or lightlike. (ii): The {\em causal past} of $p$ is the set of points that 
causally precede it: $J^-(p) := \{q : q {\rm{\; causally \; precedes \;}} p \}$. (One 
similarly defines the causal future $J^+(p)$.) But it turns out, for technical reasons, to 
be easier to work with the interior of
$p$'s past light-cone, i.e. the points connectible to $p$ by signals travelling slower than 
light. Formally, these are the points connected to $p$ by a future-directed curve whose 
tangent vector at every point is timelike: dubbed the {\em chronological past} and written 
$I^-(p)$.

Summing up this notation and jargon: Manchak's theorems will show that an ideal observer at 
$p \in M$ who knows the full metric structure of $I^-(p)$ cannot know much about the global 
structure of her spacetime, since many different spacetimes, with widely varying global 
properties, have a region isometric to $I^-(p)$. More precisely: let us say that a spacetime
$(M,g)$ is {\em observationally indistinguishable} from $(M',g')$ iff for all points $p \in 
M$, there is a point $p' \in M'$ such that $I^-(p)$ and $I^-(p')$ are isometric.  (The fact 
that this
notion is asymmetric will not matter.) Then the gist of the theorems is that almost every 
spacetime is observationally indistinguishable from another, i.e. a non-isometric spacetime.

So much for the gist. The precise theorems incorporate (i) a mild limitation, and (ii) two 
significant generalizations; as follows.\\
 \indent (i): The theorems set aside spacetimes $(M,g)$ that are {\em causally bizarre} in 
the sense that there is a point $p \in M$ such that $I^-(p) = M$. (This strong condition 
implies various causal pathologies, in particular that there are closed timelike curves.)\\
 \indent  (ii): But the theorems accommodate any further conditions you might wish to put on 
spacetimes, provided they are {\em local}, in the sense that any two spacetimes $(M,g)$ and 
$(M',g')$ that are locally isometric (i.e. any $p \in M$ is in a neighborhood $U \subset M$ 
that is isometric to a neighborhood $U' \subset M'$, and {\em vice versa}) either both 
satisfy the local condition, or both violate it. (So this generalizes the point in (3) above 
that the theorem accommodates having observations of facts about matter and radiation.)\\
\indent The theorems also veto an observer's ascertaining some global properties of her 
spacetime. Manchak lists four such properties (2011, pp. 413-414). Three are `good causal 
behaviour' properties: viz. that the spacetime be globally hyperbolic, inextendible and 
hole-free. We will not need their definitions. But the fourth property will be crucial to 
the cosmological principle (cf. Section \ref{CP1}). It is spatial isotropy: there being, at 
every spacetime point, no preferred spatial direction. Thus the theorems will veto an 
observer's ascertaining these properties: given a spacetime $(M,g)$ with any or all of these 
properties, there is an observationally indistinguishable spacetime with none of them.

Thus Manchak's theorems (2009, Theorem, p. 55; 2011, Proposition 2) say: Let $(M,g)$ be a 
spacetime that is not causally bizarre, and that satisfies any set $\Gamma$ of local 
conditions. Then there is a non-isometric spacetime $(M',g')$ such that: \\
\indent (a): $(M',g')$ satisfies $\Gamma$, but has none of the four listed global 
properties;\\
\indent (b): $(M,g)$ is observationally indistinguishable from $(M',g')$.

The proof is short, and conceptually simple. It involves a stupendous `cut-and-paste' 
construction, using (copies of) patches of the given spacetime $(M,g)$ to define the other 
one $(M',g')$---but stringing the patches out separately from one another, like shirts 
hanging on a clothes-line, so that $(M',g')$ is not isometric to $(M,g)$. The idea of the 
clothes-line construction goes back to  Malament: whose discussion also emphasizes that 
global properties which $(M,g)$ might have can fail to hold in $(M',g')$ (1977, pp. 70-74).

I have already noted that the theorems, despite their apparent formulation, allow for 
observations of matter and radiation. I should also note three other strengths.\\
\indent (i):  One might consider observations collected by a single immortal observer, or by 
an eternal dynasty of observers with each generation bequeathing its observations to the 
next. But I have here stated the theorems in terms of the past of a single point $p$, and so 
apparently made the threat of under-determination weaker. For even if $p$ is ``late'' in the 
universe, and has a vast past, might not later data break the under-determination? In fact, 
the threat cannot be allayed in this way: Malament (1977, pp. 63, 66-68). \\
\indent (ii): One might object that the theorems are too idealized. In particular, how could 
an observer ever pin down the geometry of $I^-(p)$? But there are impressive, albeit again 
idealized, theorems saying that this {\em is} possible (Ellis (1980), Ellis et al. (1985, 
especially Section 12, pp. 377-390)). The idea is that assuming general relativity and some 
technical assumptions, observations of a suitable set of `standard objects' of known size, 
mass and luminosity are necessary and sufficient to determine the geometry of $I^-(p)$.\\
\indent (iii): The definition of observational indistinguishability, and thereby Manchak's 
theorems, are in a sense stronger than they need to be, in order to show 
under-determination. For surely, under-determination  does not require that there
be {\em one} spacetime $(M',g')$ such that, for all points $p \in M$,
there is a point $p' \in M'$ with a chronological past isometric to that of $p$. Rather, it 
at most requires that for each $p \in M$, there
exists a spacetime $M'$ containing a point $p'$ whose chronological past is isometric to 
that of $p$. That is: the spacetime $M'$ can depend on the choice of $p \in M$. (Thanks to a 
referee for this point.) 

One might worry about the scientific significance of the theorems: isn't the `cut-and-paste'
construction `unphysical'? But that vague word covers several possible misgivings. One is 
that the theorems entirely concern {\em classical} general relativity, which we have every 
reason to believe fails for the extreme conditions in the very early universe and black 
holes. But this points beyond  standard cosmology---and so beyond this paper.

Within classical general relativity, the theorems are not so easily dismissed. That we 
define a spacetime model by a cut-and-paste
construction is no evidence at all that the features it exhibits are not generic among 
general relativity's models. (I will return to the idea of generic states, in Sections 
\ref{CP21} and \ref{env}.) And as to the specific four global properties that Manchak shows 
to be impossible to ascertain observationally: he reviews various authors' rationales for 
requiring one or more of these properties, concluding sceptically that none of these 
rationales are convincing (2011, Sections 5 and 6, p. 415f.). I also commend Norton's 
similarly judicious assessment (2011, especially Sections 5 and 6), and his sceptical 
conclusion that we are at a loss to justify the inductive inferences, which favour 
`reasonable' spacetimes over `gerry-mandered' observationally indistinguishable 
alternatives, that we intuitively endorse.\footnote{Agreed: there may be some constructions 
of spacetimes that we can justify condemning as contrived. Thus Magnus (2005, Section 6) 
makes a good case that we should condemn a spacetime that repeats initial data (maybe many, 
even infinitely many, times)  by disidentifying the initial data in a given spacetime.}

The next Section takes up the discussion of one of these properties, spatial isotropy, in 
the context of cosmology's standard  model. So the question will be whether, despite 
Manchak's and Norton's doubts, we can cut
down the great variety of models allowed by the Malament-Manchak theorems by appealing to a 
principle of uniformity or simplicity---the cosmological principle.

\section{The cosmological principle to the rescue?}\label{CP}
The cosmological principle (CP) is one of several principles that have been appealed to 
since the mid-twentieth century in order to make cosmological models simple enough to be 
tractable. Indeed, the CP is the strongest and probably best known of them. Roughly 
speaking, it says that at any time, the universe is approximately the same at all spatial 
positions, and in all directions. Evidently, that is a logically strong assertion, which can 
be expected to reduce the threat of under-determination---but which would also need a lot of 
justifying.

Broadly speaking, this Section describes how the CP is what is colloquially called a `lucky 
break', i.e. a piece of good fortune---so far. That is:  in Section \ref{CP1}, I will report 
that CP does greatly reduce under-determination (in physicists' jargon: constrain 
cosmological models), and that observations of many kinds suggest CP holds good, on 
sufficiently large scales, of the observable universe. But Section \ref{CP2} will urge that 
we have much less warrant for believing it of the {\em entire} universe. So, broadly 
speaking, Section \ref{CP1}'s discussion  will be upbeat, and will mention some of the great 
observational and theoretical achievements of modern cosmology. But Section \ref{CP2}'s 
discussion will be downbeat. It will bring out how much we have yet to understand; including 
about issues like what would be an appropriate probability measure over initial conditions, 
which lead into controversies about the very early universe (i.e. times logarithmically much 
earlier than one second). It also meshes with Manchak's result (Section \ref{obsind}) that 
no observer can ascertain observationally that her spacetime is everywhere spatially 
isotropic; (for spatial isotropy implies the spatial homogeneity that the CP also asserts: 
cf. Section \ref{CP12}).

Given the CP's central role in cosmology, both theoretically and observationally (and ever 
since it was formulated, in 1935), this Section will be dismayingly brief. Apart from lack 
of space, my excuse is that the CP's central role has prompted a good deal of attention to 
it, over the decades, in the philosophical literature; including comparison with other such 
principles, such as the  `Copernican principle' (e.g. Ellis 1975, pp. 252-257), and `Weyl's 
principle' (e.g. Rugh and Zinkernagel 2011). I especially recommend  recent work by  
Beisbart (2009, 2012). His (2009) focusses on the CP's breaking under-determination, and 
what follows is but a glimpse of his treatment. (Cf. also Beisbart and Jung (2006); for a 
brief overview by a {\em maestro}, cf. Ellis (2007, p. 1223-127, Section 4.2.2).

\subsection{A lucky break}\label{CP1}
The CP says that on sufficiently large scales, the universe and its material contents are 
(a) spatially homogeneous and (b) spatially isotropic about every spacetime point. Roughly 
speaking, this means that, although both spatial geometry and the matter and radiation 
content of space can change over time, there is a foliation of the spacetime into spacelike 
hypersurfaces, on any one of which: (a) at any two points, the state (of metric, matter, 
radiation etc.) is the same, and (b) at any point, all spatial directions are equivalent. 
(For precise definitions of both (a) and (b), cf. e.g. Weinberg (1972, pp. 378-381, 
409-413), Wald (1984, pp. 92-94).)

In this principle, the idea that space itself is homogeneous and isotropic is very familiar 
from Newton's physics. Beyond this, the principle expresses two ideas.

(1): The first is the material contents being required to be homogeneous and isotropic. Here 
the qualification `on sufficiently large scales' is crucial, since these material contents 
are evidently inhomogeneous and anisotropic on most observed scales. Matter is clumped 
together in bodies separated by apparently empty space. And at various scales, various 
directions are preferred; e.g. up-down, at human scales on Earth; and to/from the Sun, at 
solar-system scales. {\em A priori}, it is unclear whether local perturbations in matter 
density are small enough in size, and-or similar enough from one spatial region to another, 
that averaging over large enough length-scales will ``wash them out''. In fact, there is an 
elementary heuristic argument, based on the Newtonian approximation, that this is 
so.\footnote{In Newtonian theory, a deviation $\delta \rho$, over a length scale $L$, from 
the Universe's mean density $\rho$ is related to the perturbation $\delta \Phi$ in the 
gravitational potential $\Phi$ by: $ \delta \Phi \sim \delta \rho L^2$. Applying this to a 
pressureless FRW Universe at a late cosmic time $t$ (cf. Section \ref{CP11}), the relative 
size of the metric perturbations is given by 
\be
\delta \Phi / \Phi \; \;  \sim \;\;  (\delta \rho / \rho) (L / ct)^2 \; .
\label{john}
\ee
The CP requires that $\delta \Phi / \Phi << 1$. But eq. \ref{john} shows that this allows 
$\delta \rho / \rho$ to be $ >> 1$, provided that $L << ct$. Indeed, a macroscopic object 
e.g. a person provides a $\delta \rho / \rho  \sim 10^{30}$ over-density compared with the 
Universe's mean density, on a length scale $L \sim$ 1 metre. But at the present epoch, $ct 
\sim 10^{25}$ metres, so that $\delta \Phi / \Phi   \sim 10^{30} \times 10^{-50} \sim 
10^{-20}$. Thanks to John Barrow for explaining this to me. Cf. his (1993, pp. 121-122), 
which also explains how: (i) the temperature fluctuations $\delta T / T$ in the cosmic 
background radiation give a direct probe of metric perturbations (indeed: $\delta T / T \sim 
\delta \Phi / \Phi$) and are so tiny ($10^{-5}$) as to strongly support the CP; (ii)  eq. 
\ref{john}, together with $\delta \rho / \rho \sim L^{-2}$ (postulated by inflation), 
explains the scale-independence of metric perturbations.\label{Barrow}} 

So here, the broad physical  issue is: how large a scale must we consider in order for the 
CP to be tenable? Section \ref{CP13} will report the present-day
answer, and Section \ref{CP22} will briefly discuss the significance of inhomogeneities.

The broad philosophical issue is the nature and justification of idealizations. I will not 
pursue this, but I recommend: for general discussion, McMullin (1985), and for cosmology 
Ellis (1991, pp. 561-568; 2011 Sections 2,3, pp. 3-11) and Beisbart (2012, Section 4). For 
example, Beisbart surveys two approaches to understanding the phrase `sufficiently large 
scales' in the CP. One is coarse-graining: roughly speaking, this is taking averages of a 
quantity such as density or velocity over larger and larger volumes so that the `clumps' get 
washed out. The other is statistical: roughly speaking, this is taking the actual clumpy 
distribution of matter to be a realization of a stochastic process that is homogeneous and 
isotropic in the statistical sense that its probability distributions are invariant under 
translations and rotations. These approaches obviously raise interesting conceptual 
questions, such as how to understand probability applied to the universe---but I cannot 
pursue them.

(2): The CP's second idea is the allowance that the geometry and material contents of the 
universe can change over time. This allowance has been familiar since the mid-twentieth 
century, thanks to Hubble's discovery of the expansion of the universe, and the formulation 
of Big Bang models of its evolution by theoreticians such as Friedmann, Lemaitre and Gamow. 
But it is worth recalling how radical this allowance seemed at first. Notoriously, 
Einstein's  initial motivation in 1917 for introducing a cosmological constant $\Lambda$ 
into general relativity was to find a cosmological model that would be constant in time. 
($\Lambda$ represented a repulsive force acting at large distances which in Einstein's model 
exactly balanced the implosive tendency of gravity.) And the steady-state theory  vetoed 
this allowance by postulating a `perfect CP' which required homogeneity in time, i.e. 
constancy, as well as in space. (For the history, cf. Kragh (1996, p. 182f.), Longair (2006, 
Chapter 12).)

The CP is a `lucky break', in three ways, which I shall discuss in turn.\\
 \indent (1): It has strong and mathematically tractable, indeed elegant, consequences for 
the spacetime metric; it thereby promises to allay the threat of under-determination 
(Section \ref{CP11}).\\
\indent (2): There are impressive mathematical theorems about its relation to other 
principles, and about how observational evidence could support it (Section \ref{CP12}).\\
\indent (3): As I mentioned in the preamble to Section \ref{CP}: for the observable 
universe, it is well-confirmed (Section \ref{CP13}).

\subsubsection{Consequences}\label{CP11}
CP implies that (on sufficiently large scales) the spacetime metric ($g$ in the notation of 
Section \ref{obsind}) is everywhere of a very simple form: the now-famous  
Friedmann-Robertson-Walker (FRW) form. This means that:\\
\indent (i) there is a cosmic time-function $t$ whose simultaneity surfaces  (also known as 
`spatial slices': spacetime points with the same value of $t$) foliate the spacetime;\\
\indent (ii) the entire spacetime geometry is described by just:\\
\indent \indent (a): an integer $k = 0$ or +1 or -1, representing the curvature of the 
spatial slices ($k = 0$ means euclidean geometry, $k = +1$ means a geometry like surface of 
a sphere etc.); and \\
\indent \indent (b): a function $R(t)$ that represents the scale-factor, or `radius', of the 
universe at cosmic time $t$.\\
The evolution of $R(t)$ depends not only on $k$, but on the type of matter and radiation 
(and so on how they interact) and on the value of the cosmological constant $\Lambda$. Thus 
by choosing various types of matter and radiation, and values of $k$ and $\Lambda$, one can 
write down some dozen or two dozen simple cosmological models. Some are stationary, some 
expand forever, some expand and then re-collapse; some have a flat spatial geometry ($k = 
0$), some a curved one.\footnote{For details of this class of possible universes, in 
authoritative textbooks (deliberately different from those in footnote \ref{thermal}!), cf. 
Misner et al. (1973, Chapter 27), Hawking and Ellis (1973, pp. 124-148), Rindler (1977, 
Chapter 9), Liddle (2003, Chapters 4, 5, 15.1), Rowan-Robinson (2004, Chapter 4); or with a 
view to philosophy, Ellis (2007, pp. 1185-1191, Section 2.1). For a masterly popular 
account, cf. Barrow (2011, Chapter 3). 

There are also important subtleties about how CP yields features (i) and (ii). Thus Rugh and 
Zinkernagel (2011, Section 3) emphasize that for CP to secure the cosmic time-function in 
(i), one needs to presuppose the congruence of timelike curves representing the mean motion 
of matter (`Weyl's principle').}

Of course, in the forty years 1925-1965, before the rise of cosmology as an observational 
science with a great variety of precise observational techniques (cf. Section \ref{CP13}), 
the expanding models in this class were  regarded as giving only a broadest-brush 
description of the evolution of the universe. And of course, only for times well after any 
earliest epochs:  in particular, well after any `primeval atom' or `primeval fireball' of 
the type advocated by Lemaitre and Gamow.  Nevertheless, these models married Hubble's great 
discovery that the galaxies were receding from us---and the associated `Copernican' idea 
that this observation should be `typical', in that any observer would see galaxies 
receding---to our best theory of gravity, Einstein's general relativity. Accordingly, they 
were much investigated: and to the extent that observation allowed, their parameters (of the 
rate of expansion, the age of the universe etc.) were estimated. (For the history, cf. Kragh 
(1996, especially Chapter 6),  Longair (2008, Chapters 6, 13).) For our philosophical topic 
of under-determination, the moral is clear: for these models to be predictive enough to 
confront observations---even the meagre observations of the mid-twentieth century---their 
incorporating CP was crucial.

\subsubsection{Reasons}\label{CP12}
 So much for the CP's consequences. What reasons can be given for believing it? Section 
\ref{CP13} will report the favourable evidence,  in the straightforward inductive sense, 
rooted in observation. Here I will briefly list some theoretical reasons. By this, I mean 
theorems that in effect reduce what we would be required to observe in order
to get evidence supporting the CP.\footnote{Since these theorems are about necessary and-or 
sufficient conditions for homogeneity and-or isotropy, they are of a similar ilk to results 
about the consequences of the CP. Among those consequences, Section \ref{CP11} focussed only 
on the FRW metric. But an exposition surely should  emphasize the metric; and since these 
theorems focus on observational requirements, it is worth separating them out from the 
general idea of the CP's consequences.}

This means I will set aside other, more rationalist, grounds for believing the CP, that over 
the decades, various authors have
adduced; for example, arguments based on the `Copernican' belief that our observations 
should be `typical'. Stating and assessing such grounds and arguments has long been a 
central theme in philosophy of cosmology. But broadly speaking, recent assessments find 
these grounds and arguments unpersuasive; for example, the `Copernican typicality' argument, 
despite its appealing `modesty', is unconvincing---for we must admit there could be 
selection effects. I concur with these recent sceptical assessments, and so set aside these 
grounds and arguments. For details, I recommend Balashov (2009), Beisbart and Jung (2006, 
Sections 4, 5), Beisbart (2009, pp. 188-189, 193-199), Ellis
(1975, especially pp. 250-257), and Norton (2011).

I will first discuss purely geometrical theorems relating the homogeneity and isotropy of 
the metric to each other, and to other geometrical conditions; (then I will discuss theorems 
about physical fields). As examples, here are three theorems.\\
\indent (i): Homogeneity together with isotropy about a given point implies isotropy about 
every point on the manifold (e.g. Weinberg (1972, p. 379) \\
\indent (ii): If a spacetime is spherically symmetric (as defined by e.g. Wald (1984, p. 
120)) about every point, then it is homogeneous (Hawking and Ellis (1973, p. 135, 369). 
Combining this with (i), we infer: if a spacetime is spherically symmetric about every 
point, and isotropic about some point, then it is homogeneous and isotropic everywhere. That 
is: the CP holds.\\
\indent (iii): Isotropy about every point implies homogeneity (Weinberg (1972, p. 379)). The 
idea is that isotropy at a point means that it has a neighbourhood on which a spatial 
rotation, mapping points of the neighbourhood among themselves, preserves the metric 
structure; and if this is true at every point, then the metric is the same everywhere. (Or 
more visually: think of connecting any two points by a sequence of rotations about suitably 
chosen other points, so that isotropy applied at these other points implies homogeneity.)\\
\indent Summing up: the gist of these theorems is that we could justify the CP if we could 
justify spherical symmetry about every point, and-or isotropy about every point.

Second, there are theorems relating the metric's spatial homogeneity and isotropy to  
physical fields. Perhaps the most famous is the Ehlers-Geren-Sachs theorem (Ehlers et al. 
1968). It says, roughly speaking, that for an expanding spacetime with pressureless matter 
and radiation: if the radiation is spatially isotropic about each spacetime point, then the 
CP holds, i.e. the metric is spatially homogeneous and isotropic and therefore is FRW. This 
theorem is remarkable for two reasons.\\
\indent (i): One might argue for the theorem's antecedent by combining the `Copernican 
principle' mentioned above (that our observations should be typical) with the fact that the 
cosmic background radiation has indeed been observed to be isotropic about what amounts, for 
cosmological scales, to being `one' spacetime point: viz. Earth-1965, or with much better 
precision, Earth-2012. \\
\indent (ii): The theorem is stable. More precisely: Stoeger et al. (1995) show that almost 
isotropic radiation measurements imply that the metric is almost in the FRW form. (Cf. the 
mention of idealization in the preamble to Section \ref{CP1}.)

Finally, there are various theorems that avoid assumptions about the physical situation at 
all (or even many) spacetime points; and so whose empirical application in cosmology can be 
less reliant on the Copernican principle {\em a la} (i) just above. I thank G. Ellis for 
pointing out to me a remarkable recent theorem of this kind. Clifton et al. (2011) show that 
by observing the cosmic background radiation for an extended period of time, or by observing 
scatterings of this radiation, a single observer can in principle conclude that their past 
light-cone's metric is homogeneous and isotropic. (The theorem exploits an effect, the 
kinematic Sunyaev-Zel'dovich effect, whereby an anisotropy of the radiation, as seen by an 
object that scatters the radiation towards us, causes a distortion in the radiation's 
spectrum as seen by us.)

\subsubsection{Evidence}\label{CP13}
 I turn to assessing the observational evidence for the CP. Recall that all our 
observational evidence comes from our past light-cone: which, following cosmologists' 
practice, I have called `the observable universe'. In the notation of Section \ref{obsind}: 
all our evidence comes from $J^-$(Earth-now). So we should of course distinguish two sorts 
of evidence, or inductive inference:\\
\indent (i)  evidence that, or inference to the proposition that:  CP holds for our past 
light-cone; and \\
\indent  (ii) evidence that, or inference to the proposition that:  CP holds for the 
universe beyond our past light-cone.\\
Topic (ii) can be addressed in various ways: as we will see in Section \ref{CP2}. Here I 
just consider (i). As I said, this will be a matter of `good news': a lucky break. I 
postpone the `bad news' till Section \ref{CP2}.

In short, the good news is that if one averages (our observationally-based descriptions of) 
matter and radiation within the observable  universe, on a sufficiently large spatial scale, 
then these observations support CP. To be a bit more precise about this, we need to respect 
the facts (as we scientific realists would say!) that the observable  universe was once very 
small, that its matter inhomogeneities then were the seeds of later galaxies, and that (as I 
mentioned in (1) of Section \ref{scene}) we cannot now directly observe (by any 
electromagnetic means)
events earlier than the time of decoupling, about 400,000 years after the Big Bang. So to 
put it briefly:  we should talk about sufficiently large spatial averaging at sufficiently 
late times after the Big Bang.

Indeed, this is not just a lucky break for present-day cosmology. We are all the more lucky 
for the CP having held up very well, over eighty years, as our astronomical and cosmological 
observations grew enormously: in the type of observation we could make (especially, covering  
non-optical wavelengths: radio, microwave, infra-red, ultra-violet, X-ray), in their 
precision, and thereby in their spatiotemporal coverage or depth. To take a very simple 
example: we now know that there are tens of thousands of galaxies in each patch of the sky 
with a visual area equal to that of the full moon. Besides, over those eighty years, we have 
discovered some enormous cosmic structures: galaxies tend to be gathered in clusters 
separated by voids; and clusters  themselves tend to be gathered in 
superclusters.\footnote{Cf. e.g. Rowan-Robinson (2004, pp. 42-44, 113-119, 136-140), Longair 
(2006, pp. 253-268). For popular accounts of the discovery of clusters and superclusters,  
cf. Rowan-Robinson (1999, pp. 18-19, 52-54, 111-114), Silk (1989, pp. 192-245) and Smoot 
(1993, pp. 137-153). For full technical treatments of galaxy clustering, cf. Saslaw (1985, 
Part II; 2000).}. Such clumping on very large length-scales obviously threatens the CP. 
However---so far as we know---there is no further clumping at length-scales yet larger than 
that of superclusters. And  the observable universe is so vast as to contain thousands of 
superclusters: so that, while averaging over a typical length or volume of a supercluster, 
it still makes sense to ask whether the CP holds.

Here are some very approximate numbers.
Let us ask: how large a scale must we consider in order to see no larger structure, i.e. so 
that we can ask whether the evidence on yet larger scales supports CP? Nowadays, the  answer 
is: a vast scale, much vaster than was believed eighty years ago when the CP was formulated, 
i.e. before clusters and superclusters were found.  Namely: we  must consider scales of 
about 300 million light years or more. Yet this is still much smaller than the observable 
universe. Again let me put in numbers, albeit in a naive way.

Taking the universe to be now 10 billion = $10^{10}$ years old, our past light-cone has a 
linear spatial dimension at its base (i.e. soon after the Big Bang, but long enough after it 
so as to be well described by a standard FRW model, e.g. 400,000 years after) equal to 10 
billion = $10^{10}$ light-years. As just mentioned, a spatial box on which CP is observed to 
be valid has a side-length $3 \times 10^8$ light years. So the ratio of side-lengths, 
between the base of our past light-cone and such a spatial box, is about: $3 \times 10$. So  
consider a vast cube, built of such boxes, standing on the square base of our past 
light-cone: a cube with side-length $10^{10}$ light years. This cube consists of a total of 
$(3 \times 10)^3 = 27 \times 10^3 \approx 3 \times 10^4$ such boxes. These $3 \times 10^4$ 
boxes are a large enough set for it to make statistical sense to ask whether the CP holds.

Agreed, our past light-cone is a cone not a cube. That is:  for many boxes at the base of 
the cube, the galaxies etc. in them have now moved through the boundary  of our past 
light-cone. Once they are out, their light does not reach us; and in a universe that expands 
sufficiently rapidly, in particular one whose expansion accelerates, we would  never receive 
the light (or other signals) emitted from such boxes. So these boxes are not sampled by any 
of our evidence (perhaps ever) from our past light-cone. But no worries: there are 
nevertheless thousands of  boxes in our past light cone, so that we can ask what our 
evidence from observing these boxes suggests about the CP.

And luckily, one might say `amazingly': many different kinds of evidence suggest that the CP 
{\em does} hold across all these boxes. (Agreed: (i) not as many as $3 \times 10^4$ boxes; 
cf. the concession in the previous paragraph; and (ii) only approximately; cf. the 
discussion of idealization at the start of Section \ref{CP1}.) The most famous evidence is 
of course the isotropy of the cosmic background radiation; (cf. (i) in footnote 
\ref{Barrow}). But there are several other, largely independent, kinds of evidence, about 
optical, radio and X-ray sources. (For this evidence, Beisbart (2009, p. 186, 199) cites 
Kahav 2001, Sarkar et al (2009); I am also grateful to W. Saslaw for pointing me to the most 
extensive recent discussion: Yang and Saslaw (2011).)

\subsection{Beyond the evidence: through a glass darkly}\label{CP2}
Without wanting to take Saint Paul's name in vain, I cannot resist stealing his metaphor to 
describe cosmology's epistemic situation, in endeavouring to go beyond the evidence for the 
CP from within the past light cone of Earth-now; in Section \ref{obsind}'s notation: the 
evidence from within $J^-$(Earth-now).

In this Subsection, I will first  discuss some strategies for going beyond our causal past 
(Section \ref{CP21}). Then I will report doubts that have (rightly!) been raised about 
whether the CP holds. These doubts are based on considering inhomogeneities; and they 
question whether spacetime is homogeneous and isotropic (in the relevant sense, i.e. as 
regards matter, radiation and geometry, on sufficiently large scales) not just beyond, but 
also within, our past light-cone (Section \ref{CP22}).\footnote{So agreed: I could have 
reported these doubts in Section \ref{CP1}, instead of in this Subsection. But I think that 
would have been less clear. Anyway, placing them here means that the distinction between 
Section \ref{CP1} and this Subsection is, not so much between our causal past and the rest 
of spacetime, as between the upbeat message in favour of the CP and the downbeat message 
against it---as I announced in Section \ref{CP}'s preamble.}

\subsubsection{Strategies for justifying the CP}\label{CP21}
I said in Section \ref{CP13} that the question whether the CP
holds for the universe beyond our past light-cone can be addressed in various ways. As I see 
matters, there are three main ways. I will just note the first two strategies, and say more 
about the third strategy.

(1): {\em Copernican?}: First, there is the approach I noted at the start of Section 
\ref{CP12}: over the decades, various authors have adduced various rationalist grounds for 
believing the CP holds good for the whole universe. But I fully concur with the recent 
sceptical assessments of these grounds (by Beisbart and others, cited there). So I find this 
approach unpromising, and I would even say that it is over-emphasized in the philosophy of 
cosmology:  so  I  set it aside.

(2): {\em Inference?}: The second way of addressing the issue is also somewhat `aprioristic' 
(and so seems natural to philosophers). It is in play, but rarely explicit, in various 
discussions of the CP. Namely, one considers the various choices of spacetime region, beyond 
the observable universe $J^-$(Earth-now), that one might consider evidence about, or 
inference to; and then envisages that the evidence or inference will have different 
strengths or plausibilities for different regions. As one would expect, the considerations 
in play usually vary from one cosmological model (or family of models) to another, 
prohibiting a general discussion. But by way of illustration, here
are three obvious specifications of spacetime regions one should consider:\\
\indent (i): the causal past of our cosmic time-slice, i.e. $J^-(t $= now) (for a model that 
has a cosmic time-function);\\
\indent (ii): the causal future of the observable  universe, i.e. $J^+(J^-$(Earth-now));\\
\indent (iii): the entire spacetime manifold, i.e. all spacetime points that are 
topologically connected to Earth-now.\footnote{Neither of regions (i) and (ii) includes the 
other; but of course region (iii) includes them both. And of course, other definitions are 
possible, and sometimes relevant.}

Mention of such regions prompts two salutary reminders about general methodology. First: in 
assessing evidence and making inferences, we must bear in mind our claims' implications for 
events in the unobservable regions, and conversely, how those events could constrain what we 
observe. (For a lucid introduction to these issues, cf. Ellis (1975)).

Second: although I noted in Section \ref{CP13} how vast the observable universe is---how 
much vaster are regions such as (i) to (iii)! Just how vast they are is of course unknown, 
not least because estimates depend on the details of how we extend the standard cosmological 
model to times much earlier (logarithmically) than one second---extensions which are likely 
to long remain controversial (cf. Section \ref{env}). But to give an uncontroversial 
example, based on a `late departure' from our past light cone: consider matter or radiation 
that left our past light cone at about the time of decoupling---as it might be, a cosmic 
background photon. That photon is estimated to now be about 40 billion light-years from us 
(the distance being measured along the cosmic time-slice  $t$ = now). Vast indeed. This 
vastness emphasizes how risky is the induction from CP's holding in the observable  universe 
(presumed on account of the evidence in Section \ref{CP13})  to its holding throughout any 
such region as (i), (ii) or (iii). To put the point very simply, in terms of enumerative 
induction over spacetime regions: the observable universe is such a small fraction of such 
regions, that it is risky to claim it is a fair sample.\footnote{For more discussion of this 
risky induction, cf. Beisbart (2009, pp. 184-5, 199-201).}

This riskiness is undoubtedly part of people's motivation in their search for general 
theoretical grounds for the CP holding throughout spacetime. I have set aside various 
rationalistic grounds, since I am sceptical of them. But there is a third way ...

(3): {\em To be expected?}: Third: Suppose we are given a cosmological theory, rather than a 
specific model, so that we envisage a range of possible initial conditions: where `initial' 
need not mean `at the Big Bang', but may mean only something like `at times earlier than one 
second'. (So the theory could be some quantum field theory on a highly curved spacetime.) 
And suppose we could show that any, or any typical, initial condition at times earlier than 
one second evolves into a later state that is homogeneous and isotropic---approximately, 
i.e. on the sufficiently large scales that we now observe the CP holding. According to most 
physicists' and philosophers' views of explanation, such a demonstration would count as an 
explanation of the CP: indeed a satisfying one, that would be a great credit of the theory.

Note that in this situation, two familiar general ideas  are in play: one philosophical, and 
one physical.\\
\indent (i): The philosophical idea is that to explain is to show that the {\em explanandum} 
was to be expected: an idea with a tradition stretching from Aristotle to contemporary 
deductive-nomological and causal accounts of explanation.\\
\indent (ii): The physical idea is a bit more specific, though also independent of theory. 
It is that initial conditions are a matter of mere happenstance, in some sense that a 
theory's laws (in particular: equations of motion) are not.\footnote{Two remarks. (i): This 
idea is of course compatible with the Humean view that laws are metaphysically and-or 
logically contingent: though the required stronger sense of `mere happenstance' is rarely 
spelt out. (ii): I have stated the idea in terms of initial conditions: as is usual, and as 
suits our cosmological discussion. But one can, and some authors do, state the same idea for 
final and-or boundary conditions.} So to show that a final state of affairs follows by the 
laws from an initial condition that is known, or reasonably believed, to have held, 
is---though informative, and perhaps even deserving of the name `explanation'---not wholly 
satisfying. For if the initial condition invoked seems a matter of mere happenstance, then 
the demonstration in effect prompts the question: why did that initial condition hold?\\
\indent Putting the two ideas together, one naturally concludes that an explanation (or at 
least: a satisfying explanation) of a final state of affairs must show, not just that it 
follows by the laws from an initial condition that is known, or reasonably believed, to have 
held, but also that it follows from any, or any typical, initial condition---and so was `to 
be expected' in a stronger sense, independent of belief in a single initial condition.

Obviously, we are here at the threshold of several broad issues, which this paper can only 
mention, not pursue; though Section \ref{env} will add a little.\\
\indent (a): One issue is the history and philosophy of concepts of explanation, especially 
in relation to features of the universe as a whole. I recommend the work of McMullin (e.g. 
1993, 2005, 2007): he calls the idea that  typical (or even all) initial conditions lead to 
a final state with the feature of interest a `Principle of Indifference'.\\
\indent (b): A second issue is the idea of happenstance: can one make convincing precise 
sense of the idea that some particular propositions, such as initial conditions, are 
contingent in a stronger sense than are (according to we Humeans) the laws of nature? Note 
that one faces this issue, even if one is sceptical (as I am) that there is a single concept 
of explanation (or even: scientific explanation).\\
\indent (c): Third: the CP, i.e. homogeneity and isotropy, is by no means the only feature 
of the `late universe' for which the topic of the scope and limits of its explanation by 
invoking earlier states has been much discussed. Another main example is the direction of 
time, and its explanation in terms of an early `smooth' equilibrium state.\footnote{For this 
example, recent philosophical discussion  emphasizing the topic of explanation includes the 
debate between Callender (2004) and Price (2004) (the former downbeat, the latter upbeat, 
about how much one can ask of an explanation). For philosophical surveys that admirably 
clarify the physics involved, I recommend Callender (2011), Wallace (2010, 2011).} Section 
\ref{env} will mention two other examples.\\
\indent (d): Finally, a fourth issue arises from the fact that we can hardly expect to 
demonstrate that according to our cosmological theory, every initial condition evolves to a 
later state with the desired feature. So we must face the issue: what do we mean by a 
`typical' initial condition? We can of course allow the answer to depend on the theory. In 
one theory, `typical' might mean `an open set in the topological space of initial 
conditions'; in another theory, it might mean `with probability greater than 0.5 (or 0.9) 
according to some natural probability measure on the space of initial conditions'. But we 
must expect criticism and controversy about the physical, or explanatory, significance of 
the topology or probability measure we have invoked.

Returning to our concern with justifying the CP: I note that two research programmes---a 
past one, and a current one---have pursued this third (`To be expected') strategy for 
justifying the CP for the whole universe. The past programme is Misner's chaotic cosmology 
programme, the current one inflationary cosmology. Space prevents me going into any details; 
(though Section \ref{env} will give a general summary of inflation). I recommend the 
following, which also emphasize philosophical issues: Barrow (1993, pp. 123-127), Barrow and 
Tipler (1986, pp. 420-430), Beisbart (2009, Section 5.1, pp. 190-193), Earman (1995, pp. 
144-148, 152-155), Ellis (1999, pp. 706-713; 1999a, pp. 57-61; 2007, Sections 2.6 and 6.3, 
pp. 1208-1210, 1237-1238) and McMullin (2005, pp. 601-604; 2007, pp. 72-74); and for a 
popular account, Barrow (2011, pp. 161-168, 198-217).\\
\indent So here it must suffice to sound two sceptical notes about these programmes, 
especially in relation to justifying the CP.\\
\indent (i): Misner's  programme is generally regarded as having failed, for a variety of 
reasons (reviewed by e.g. Barrow (1993, pp. 123-124) and Earman (1995, pp. 144-145, 148)).\\
\indent (ii): As to the inflationary programme, Section \ref{env3} will mention various 
reasons for scepticism about it. Here, I just follow Beisbart (2009, p. 192) in pointing out 
that although inflation {\em prima facie} seems to produce homogeneity by vastly stretching 
out inhomogeneities (cf. Section \ref{env1} for its solution to the flatness problem), this 
cannot really amount to justifying the CP. For inflationary cosmology envisages many 
stupendously distant regions of the entire universe that are typically---absent some 
fine-tuning of initial conditions, which would flout the spirit of the programme---{\em not} 
approximately homogeneous.

\subsubsection{Inhomogeneities: doubts about the CP}\label{CP22}
I close this Section by mentioning some doubts, based on considering inhomogeneities, about 
whether the CP holds not just beyond, but also within, our past light-cone. I will follow 
Ellis's recent survey of the issues (2011); (I also recommend the papers in the same issue 
which he cites).

So the over-arching question is whether the CP, or the FRW models it implies, is a justified 
large-scale description of the universe, despite the fact that it (and even its observable 
part) is evidently inhomogeneous in many ways and on many scales. Ellis distinguishes three 
main issues, and discusses them {\em seriatim}:\\
\indent (1): inhomogeneities may affect the large-scale dynamics;\\
\indent (2): inhomogeneities affect light propagation, and so may affect cosmological 
observations;\\
\indent (3): maybe the universe is after all not spatially homogeneous on the largest scales 
and is better represented at late times by a Lemaitre-Tolman-Bondi spherically symmetric 
inhomogeneous model.\\
I shall set aside (2), but say a few words to explain (1) and (3).

(1): {\em Inhomogeneities and dynamics}:--- The basic point is that in any dynamical system, 
coarse-graining in the sense of partitioning the space of states can fail to `mesh' with the 
dynamics. That is: if $s_1, s_2$ are two states that the coarse-graining assigns to the same 
cell of the partition of the state-space (usually because $s_1, s_2$ match in their values 
for a select subset of quantities, usually called `macro-quantities'), the later states into 
which $s_1, s_2$ evolve may well {\em not} be in the same cell. So at the level of the cells 
(the macro-quantities), there is no well-defined dynamics. Or in other words: the given 
dynamics is a well-defined `micro-dynamics', but the coarse-graining does not properly 
define a `macro-dynamics'.  \\
\indent This point is often made: sometimes (as by Ellis here) in the jargon that the 
dynamics does not {\em commute} with the coarse-graining. This matches the algebraic jargon, 
that a square diagram of functions (with arrows for time-evolution and for coarse-graining) 
can fail to be `a commuting diagram'. Besides, even if there is commutation, the 
macro-dynamics can look very different from the micro-dynamics. For introductory discussion 
and examples, cf. Butterfield (2011, Section 3, pp. 107-110), Ellis (2011b, Section 
3.4.2).\\
\indent Returning to our specific topic: here of course, we are given, {\em not} the 
micro-dynamics,  which will be myriadly complex---even if, as is often done both in theory 
and in simulations, each large mass, say each galaxy, is treated as a point-particle. 
Rather, we are given a putative macro-dynamics, viz. the equations of general relativity, in 
a form that describes an expanding FRW model (and in order to describe time-evolution, cast 
in a `3+1' form, rather than as Einstein's original field equations). And our question is 
whether this macro-dynamics is induced by a reasonable coarse-graining procedure on the 
unknown, myriadly complex, micro-dynamics. If we can be confident that it is thus induced, 
then the macro-dynamics, i.e. the FRW model, is indeed a justified large-scale 
description.\\
\indent  But evidently, this is a difficult question to answer! Indeed, even if we knew the 
complex micro-dynamics and coarse-graining procedure in full detail, it would probably be 
difficult to answer---for computational reasons. So it is perhaps unsurprising that Ellis 
goes on to survey no less than four substantially different, and still open, approaches to 
answering it; (2011, Section 2.3-2.5, pp. 4-9; cf. also Clarkson et al. 2011).\\
\indent I should also mention what amounts to the `converse' topic or question; (if only to 
distinguish it from our topic). It is often said that though the universe is expanding, 
gravitationally bound systems like the Earth or solar system are not expanding. In terms of 
the familiar pedagogic metaphor that the universe is like an inflating rubber balloon: such 
systems are like pennies glued to the balloon, with the glue being stiff enough that the 
rubber below a penny does not expand. Indeed, a vivid metaphor. But it prompts the 
questions: in the real universe, how large are the pennies---planets? solar systems? And 
what is the glue: i.e. the mechanism by which small enough systems are decoupled from the 
expansion? Thus this `converse' topic is: the influence of the cosmological expansion on 
local dynamics. A wonderful topic, with a long history---and a masterly recent review by  
Carrera and Giulini (2010).

(3): {\em Inhomogeneous models}:--- By and large, this paper has hitherto celebrated the 
achievements of the standard cosmological model, perhaps as modified by inflation (cf. 
Section \ref{env}); though (i) I have duly admitted the existence of respectable rivals ((b) 
at the end of Section \ref{scene}), and (ii) this Subsection (Section \ref{CP2}) has had the 
downbeat message that we cannot now adequately justify the CP beyond the observable 
universe. In particular, I  said in footnote 3 that the $\Lambda$CDM model is the best fit 
of the standard model to all cosmological observations we have made.\\
\indent But now the cat is out of the bag! The point here is that the $\Lambda$CDM model 
being the best fit of the standard model does not imply, of course, that it is the {\em 
unique} best fit model. And there is considerable evidence that the observations we have 
made so far can be equally well fitted by Lemaitre-Tolman-Bondi spherically symmetric 
inhomogeneous models---without, one might add, the all-too-conjectural dark energy of the 
$\Lambda$CDM model. Indeed, this sort of rival to the standard model has a long history: 
apart from Ellis' recent introduction (2011, Section 4, pp. 11-16), cf. e.g. MacCallum 
(1979), Ellis (1991, pp. 563-564; 1999a, p.A60).\\
\indent But again, I cannot pursue this topic. I just note that from a broad viewpoint, it 
represents `business as usual', scientifically and philosophically. Scientifically,  it may 
be possible, within the next ten to twenty years, to make observations that discriminate 
between the models. And meanwhile, philosophically: we have again illustrated the theme of 
cosmological under-determination: and more specifically, Section \ref{CP2}'s downbeat 
message that we cannot now adequately justify the CP.

\section{Pressing the question `Why?'}\label{env}
When one is given a description of anything, it is tempting, and perhaps inevitable, to 
press the question `Why is that so?'. Besides, on most accounts of scientific explanation 
that distinguish description and explanation,  one can legitimately press the `Why?' 
question even when one is given an
explanation. In particular, one can press it when given what this paper has celebrated: the 
established description of the observable universe from the time of primordial 
nucleosynthesis. Asking {\em why the CP should hold} (of the observable universe, or of the 
entire universe) forms one way of pressing the `Why?' question in this context. 

I will end this paper by  briefly raising two other ways of pressing the question. Agreed: 
they lead to physical issues about times much earlier (logarithmically!) than one second 
after the Big Bang, and so to controversy: which in Section \ref{prosp}, I placed outside 
this paper's scope. But they relate so closely to justifying or explaining why the CP holds 
that some discussion, albeit brief, is in order.

Thirty years ago, these questions were dubbed `the flatness problem' and `the horizon 
problem'. Here `problem'  means `problem for the then-current standard cosmological model 
(so incorporating the CP and its FRW metric)'. In brief: the problems were that this model 
stated that, but did not explain {\em why}:\\
\indent (i) The universe at times later than about one second has an approximately flat 
(i.e. euclidean) spatial geometry: (`the flatness problem'; Section \ref{env1}).\\
\indent (ii) The photons in the cosmic background radiation that come to us along different 
directions in the sky have the same spectrum, even for directions different enough so that, 
according to this model, the regions on the last scattering surface, from which the photons 
were emitted 400,000 years after the Big Bang, at a temperature of about 4000 K, could have 
had no previous causal contact with one another. This lack of contact means that there could 
not have earlier been any equilibration process by which the regions attained the same 
temperature: (`the horizon problem'; Section \ref{env2}).

These explanatory lacunae were among the motivations for inflationary cosmological 
models---which apparently {\em did} explain (i) and (ii): namely, as arising from a 
conjectured exponential expansion during very early times, at about $10^{-35}$ seconds (e.g. 
Guth (1981)). The idea is that, given this exponential expansion:\\
 \indent (i): our entire universe arose from inflating a minuscule very early patch of 
spacetime, thus  explaining its current approximate flatness; \\
 \indent (ii): the emission events for any two cosmic background photons, even two that come 
to us along opposite directions in the sky, {\em do} have a common past, i.e. their past 
light cones intersect; so that there could have been an earlier equilibration process making 
the temperatures of the emitting regions equal.\\
(Of course, the merits of inflationary  models go well beyond these two explanations: in 
particular, they include predictions about the fluctuations in the photons' temperature, 
which have been confirmed by the COBE, WMAP and Planck projects---but I cannot discuss these 
here.)

Thus we return to the idea of explaining a feature of the `late' universe by invoking 
conjectural earlier states, and to issues about the scope and limits of such explanations 
(cf. the third strategy, (3), of Section \ref{CP21}). In spelling out some details, we will  
glimpse (in Section \ref{env3}) how inflationary cosmology's answer to these problems has 
led to proposals
for, and  much controversy about: both a multiverse ontology, and the idea of explaining 
features of the universe as arising from a selection effect (so-called `anthropic 
explanations').

\subsection{The flatness problem}\label{env1}
Today's cosmologists favour inflationary cosmological models over the original standard 
models, on the grounds that they explain the value of the density parameter $\Omega$. In 
Newtonian terms, $\Omega$ is, roughly speaking, the ratio of the universe's gravitational 
potential energy
to its kinetic energy. In general relativity, it is the ratio of the density of matter, 
radiation etc. to the density that would make the universe spatially flat, i.e. make the 
instantaneous spatial slices have asymptotically a euclidean geometry. So $\Omega$ is a pure 
dimensionless number.  Observations show it to be very close to 1, i.e. its theoretically 
privileged value which makes for a flat universe. (For these details, cf. e.g. Liddle (2003, 
pp. 20, 48, 52-55, 132); $\Omega = 1$ corresponds to $k = 0$ in (ii: a) of Section 
\ref{CP11}.)

The original standard models treated the value of $\Omega$ as a brute fact, in the sense of 
being a fact that is set aside from the quest for explanation. Agreed, it is no doubt a 
matter of judgment which facts it is legitimate to treat as brute; and this is a judgment 
that often varies from one context or episode of enquiry to another, and which may be 
contentious or even subjective. An oft-cited historical example of this sort of variation is 
Kepler's invoking a nested sequence of the Platonic solids, snugly fitting into each other 
like Chinese boxes, to explain the relative sizes of the orbits of the planets: but this is 
a fact which Newtonian celestial mechanics later came to treat as brute.\footnote{This 
example also illustrates how  strongly, though unconsciously, education inculcates a sense 
of which facts should be treated as brute: anyone learning Newtonian mechanics, yet unaware 
of Kepler's struggles, would never think to ask why the planets' orbits are the sizes that 
they are. We are taught to treat those sizes as `matters of mere happenstance'; (cf. the 
discussion in (3) (ii) of Section \ref{CP21}).} So for all that I have so far said, one 
might well conclude that the original Big Bang/standard models treating $\Omega$ as a brute 
fact was no discredit to them.

But it seems unsatisfying to have nothing to say about why a parameter is close to its 
theoretically privileged value. And it is all the more worrying if being close to this value 
now requires being much closer to it, in the past---as happens in this case: for $\Omega$ to 
be about
1 today required that it be extraordinarily close to 1 very soon after the Big Bang. For 
example: according to a simple model, assuming merely that $|\Omega(\rm{now}) - 1| < 0.5$ 
implies that $\Omega$  has to be  $1 \pm 10^{-18}$ one second after the Big Bang, and $1 \pm 
10^{-30}$ at $10^{-12}$ seconds after it. (This is because in this model $|\Omega(t) - 1| 
\propto t$, and the present age of the universe is about $4 \times 10^{17}$ seconds; cf. 
e.g. Liddle (2003, p. 100).) I think the natural reaction to so exquisite a tuning of a 
dimensionless number to its privileged value is that it {\em calls out} for explanation: we 
should not treat the observed value as a brute fact.

In any case, this has been the reaction of cosmologists---at least, {\em after} the 
inflationary models were devised, so that one had a putative explanation to assess and 
perhaps endorse. Besides, the basic idea of this explanation was attractively simple, and 
the same for the various inflationary models. Roughly speaking, the idea is that any 
manifold `looks flat' on a sufficiently small scale. More formally: about any point in a 
Riemannian manifold, there is a neighbourhood that is sufficiently small that its metric 
structure is as close as you please to being euclidean. Transferring this idea to the 
context of an expanding spacetime, we infer that if  our entire universe arose from a 
suitably fast, say exponential, expansion of a minuscule very early patch of spacetime, its 
spatial slices would have, after this expansion, an approximately euclidean geometry. A bit 
more specifically, in terms of inflationary cosmology: $\Omega$ will be driven very close to 
1 by the end of the inflation period. (For some details, cf.  Lawrie (1990, pp. 333-334), 
Liddle (2003, p. 104-107).)

\subsection{The horizon problem}\label{env2}
As emphasized above (especially Section \ref{CP13}), the cosmic background radiation, dating 
from the decoupling time $t_{d}$ (400,000 years after the Big Bang), is to a very good 
approximation isotropic.  This means that all the emitting regions were at about the same 
temperature. But for two directions in space with a sufficient angular separation, the two 
past events at the time $t_{d}$ that lie along those directions, i.e. the two events of 
emitting a photon that we receive along that direction, had---according to the standard Big 
Bang models---no common causal past. That is, their past light cones did not intersect. 
Calling the events $p, q \in M$, we have: $J^-(p) \cap J^-(q) = \emptyset$. This causal 
separation means that no equilibration process (or any other sort of interaction) could have 
occurred earlier, so as to establish the emitting regions' almost equal temperatures.

Similarly to the flatness problem, one might at first think that this is curious, but {\em 
not} an intellectual problem. That is, one might think that the standard models' thus 
needing an early state (at least as early as $t_{d}$) to be special, in that it encodes 
these equal temperatures, is no discredit to them. Why should we hope that the features of 
early states can be explained in some strong sense, for example by exhibiting them as 
generic? Of course, the notions of `explanation' and `generic' in play here would need to be 
clarified: cf.  Section \ref{CP21}'s discussion of its third strategy, (3). But the idea 
will be: the most we can reasonably hope for is that they are retrodicted by our theory's 
description of later states (which we will no doubt have to describe collectively, rather 
than individually, to get enough information for a retrodiction): `the early states were 
what led to {\em this}---end of story'.\footnote{This irenic response can be filled out 
somewhat, using our notation $J^-$ for causal pasts, in the context of a 
future-deterministic relativistic theory---such as we would expect a classical cosmology to 
be. (This point is made by Earman (1995, pp. 139-140), in the course of his masterly 
discussion (ibid., Chapter 5) of the horizon problem.) In such a theory, the state on a 
hypersurface $\Sigma$ that extends across a
spacetime region $R$ determines the state throughout the {\em future domain of dependence} 
$D^+(\Sigma
\cap R)$. (The future domain of dependence $D^+(Q)$ of any region $Q$ is defined as the set 
of spacetime points through which any past-directed causal (everywhere timelike or 
lightlike) curve must intersect Q. Intuitively, it is the set of points on whom any causal 
influence must first register on $Q$.)  So if we (i) pick the region $R$ to be $J^-(p) \cup 
J^-(q)$, and (ii) take $\Sigma$ to extend across $R$ prior to the `summit' of $J^-(p) \cap 
J^-(q)$, and  (iii) consider as an initial condition (or as a  premise of a 
deductive-nomological explanation) the state on $\Sigma \cap (J^-(p) \cup J^-(q))$: then we 
can surely give a prediction, and so surely an explanation, of the states at $p$ and at $q$. 
As the song says: `who could ask for anything more?'}

But I think that the quantitative details show that there is indeed an intellectual problem, 
or at least `embarrassment'. (Again, I think this is analogous to the flatness problem: 
there, it was the quantitative exquisiteness of the tuning of $\Omega$ that called out for 
explanation.) I have in mind not just the nearness of the equality of the temperatures,  for 
all directions in the sky: namely, equality to about one part in $10^5$. Also, and perhaps 
more striking: the directions in the sky for which, according to the standard models, the 
photon emission-events have no common causal past, can be very close. You might expect that 
they need to be approximately perpendicular, or even opposite, to each other, for there to 
be the `embarrassment' of same temperature with no possible explanation (by the standard 
models) invoking previous interaction. But no: the directions can be separated by as little 
as one or two degrees; (cf. e.g. Liddle (2003, pp. 102, 109), Lawrie (1990, p. 326-7). Since 
there are 360 degrees in a circle, this means that according to the standard models, the 
last scattering surface is composed of tens of thousands of regions, any pair of which has 
no common causal past---yet all `embarrassingly' at the same temperature, to about one part 
in $10^5$.

Besides, the number of causally disconnected regions grows embarrassingly steeply, as one 
runs the standard model further back in time than $t_d$. Barrow (1993, p. 126; 2011, p. 205) 
gives a simple estimate: at $t = 10^{-35}$ seconds, the observable universe would have been 
about 1 cm across, and a causally connected region would have been about $10^{-24}$ cm 
across. So, thinking naively of these regions as forming a cube: there would then have been 
$(10^{24})^3 = 10^{72}$ such regions, all mutually causally disconnected.

In any case, again analogously to the flatness problem: the cosmologists' verdict has been 
that there is a problem---at least, {\em after} the inflationary models were devised, so 
that one had a putative explanation to assess and perhaps endorse. And again, the basic idea 
of this explanation was attractively simple, and the same for the various models: the 
inflation period  implies that the events $p,q$ have a common causal past---even for two 
opposite directions in the sky. That is: $J^-(p) \cap J^-(q) \neq \emptyset$, so that there 
could have been an earlier process of equilibration of the desired kind.

Finally, I should mention a connection with a worthy perennial of the philosophy of 
probability and causation, as well as of quantum non-locality: Reichenbach's {\em Principle 
of the Common Cause} (PCC: 1956, Section 19). Roughly speaking, the PCC says that if two 
events are correlated, but neither causes the other, then in their common causal past there 
is a third event, conditional on which they are probabilistically independent (called the 
`common cause' or `screener-off') . Notwithstanding this abstract formulation, countless 
examples in everyday life and the sciences exemplify the principle, making it intuitively 
plausible. And clearly, the `embarrassment' that constitutes the horizon problem  echoes 
this plausibility: since the two emission-events' temperatures are so well correlated, 
surely some third event in their common causal past influenced them 
both.\footnote{Hofer-Szabo et al. (2013) is a definitive monograph on the PCC. My own view 
of its role as a motivation for the locality assumptions in Bell's theorem is in Butterfield 
(2007). Here it must suffice to note a contrast between quantum non-locality and 
inflationary cosmology. (1): As Hofer-Szabo et al. emphasize, Bell's theorem needs to 
assume, not only that the two events in each of several pairs have a common cause (in the 
sense of rendering the two events probabilistically independent), but also that one and the 
same event is the common cause for the several pairs (and so is called a `common common 
cause'): an assumption which Nature apparently violates. (2): In inflationary cosmology, the 
physical situation is of course both much more complicated and much more conjectural. But 
the idea is that the photon-temperature correlations observed between many, if not all, 
pairs of directions in the sky are screened off by a single, suitably extended, event or 
process of equilibration in the pairs' common past. So using Hofer-Szabo et al.'s jargon, 
the idea is that there is a common common cause: an assertion which---if we judge by the 
recent evidence favouring inflation---Nature obeys.   Finally, I should also note that the 
{\em maestro} Earman  takes a broadly sceptical view of the significance of the PCC, and 
indeed of the `embarrassment'; (1995, pp. 135-140, 142-146, 156; Earman and Mosterin 
(1999)).}

\subsection{Questions beyond}\label{env3}
So much by way of motivating the idea of inflation. So far, the broad methodological 
situation seems as described in the third approach, (3), of Section \ref{CP21}. Namely: one 
buys a satisfying explanation of a `late' feature of the universe, by paying the price of a 
speculative piece of physics for very early times.

But this is not to suggest that the situation is simple, or settled. Apart from the broad 
issues about explanation etc. mentioned in (3) of Section \ref{CP21}, there are many 
physical details relevant to the methodological issues: for example, about these models 
assuming a wholly conjectural inflaton field, about there being embarrassingly many such 
inflation models, and about how these models are meant to secure that the inflation period 
ends. (These issues are pressed by, e.g. Ellis (1999a, p. A59;  2007, Section 5, pp. 
1232-1234), Earman and Mosterin (1999).)

Besides, some of these details raise much wilder ideas, in particular the multiverse: which, 
by way of conclusion, I briefly report. (For details, cf, e.g. Albrecht (2004), Linde (1987, 
1990, 2004).)\\
\indent (1): First, the idea of inflation leads to require models in which, during the 
inflation period, countless spacetime regions branch off and themselves inflate to yield a 
`baby universe'. Since what we have so far called `the entire universe' (or in other words: 
`a standard cosmological model') would correspond to one such region (`our region'), the 
totality of regions is dubbed `the multiverse'. (But `many-headed Hydra' would be a more 
vivid name.)\\
\indent (2): Besides (though I have not yet admitted it): inflationary models are quantum 
theoretic, so that the events of a region branching off, and of its inflating, and the 
details of how the branching and expansion occur, are all stochastic events, with 
probabilities given by quantum theory.\\
\indent (3): Furthermore, the idea of inflation requires models in which, generically, the 
stochastic process of growing branches goes on forever. Hence the slogan that `inflation is 
generically eternal'.\\
\indent (4):  And finally: some inflationary models assume that the prevailing laws of 
physics, in particular the values of physical constants (such as the charge of the electron, 
or the ratio of the strengths of the gravitational and electromagnetic forces), vary across 
the different regions. To take a simple example: consider a region whose state just after 
its inflation period matches our own region's state at the corresponding time, i.e. just 
after our inflation period, except that this region has a  gravitational force a little 
stronger than ours. That force would soon halt the region's post-inflation expansion, and 
lead to collapse, called a `Big Crunch' on analogy with `Big Bang'. (For numerical estimates 
for `a little stronger' and `soon', cf. e.g. Dicke and Peebles (1979, p. 514).)

To sum up: the totality of all these spacetime regions is a stupendous, and in general 
infinite, probabilistic branching structure. The regions are linked together by narrow 
throats, rather like the elongated bubbles in the balloon sculptures of dogs or what-not 
made by magicians and street performers. And according to some models, the laws of physics, 
in particular the values of physical constants, vary across the regions. Obviously,  we here 
glimpse many
undoubtedly wild questions: both physical and philosophical---about what should count as 
explaining initial conditions or the values of constants, about selection effects, 
ontological parsimony etc. Opinion is strongly divided about such questions, and I must 
postpone discussion of them to another place.\footnote{For {\em maestri} being sceptical 
about inflation, I recommend:  Ellis (1999, p. 706-707; 1999a, pp. A59, A64-65; 2007, 
Section 5, pp. 1232-1234), Earman (1995, pp. 149-159), Earman  and Mosterin  (1999), Penrose 
(2004, pp. 735-757, Chapters
28.1-28.5) and Steinhardt (2011).  For philosophical discussion of the multiverse and 
selection effects, I recommend: Ellis (1999, pp. 696-700, 707-712; 2011a), Smeenk (2012, 
Sections 6 to 8), Wilzcek (2007) and the other papers in Carr (2007), and Zinkernagel (2011, 
Section 4.1).}\\ \\

{\em Acknowledgments}: This paper is dedicated to the memory of Ernan McMullin, whose 
contributions to the philosophy of science were so wise, insightful and scholarly; and whose 
energetic and generous spirit enhanced so many lives. For
correspondence and comments on previous versions, I am very grateful to the editor and two 
anonymous referees; and to Nazim Bouatta, Claus Beisbart, George Ellis, John Manchak, John 
Norton, Brian Pitts, Chris Smeenk, Arianne Shahvisi, Bill Saslaw, David Sloan, Bill Stoeger,
Henrik Zinkernagel, Lena Zuchowski. I also thank John Barrow, George Ellis and Henrik 
Zinkernagel for encouragement, and for the inspiration of their writings. Work on this paper 
was supported  by a grant from the Templeton World Charity Foundation, which I gratefully 
acknowledge.\\

\section{References}
Albrecht, A. (2004), `Cosmic inflation and the arrow of time', in J. Barrow, P. Davies and 
C. Harper eds., {\em Science and Ultimate Reality: quantum theory, cosmology and 
complexity}, Cambridge  University Press, pp. 363-401.

Balashov, Y. (1994), `Uniformitarianism in cosmology: background and philosophical 
implications of the steady-state theory', {\em Studies in History and Philosophy of Science} 
{\bf 25}, number 6, pp. 933-958.

Balashov, Y. (2009), `A cognizable universe: transcendental arguments in physical 
cosmology', in ed. M. Bitbol et al. {\em Constituting Objectivity}, Springer, pp. 269-278.

Barrow, J. (1993), `Unprincipled cosmology',  {\em Quarterly Journal of the Royal 
Astronomical Society} {\bf 34}, pp. 117-134.

Barrow, J. (2011), {\em The Book of Universes}, London: Bodley Head.

Barrow, J. and Tipler, F. (1988), {\em The Anthropic Cosmological Principle}, Oxford 
University Press.

Beisbart, C. (2009), `Can we justifiably assume the Cosmological Principle in order to break 
model under-determination in cosmology?', {\em Journal of General Philosophy of Science} 
{\bf 40}, pp. 175-205.

Beisbart, C. (2012), `The many faces of the Cosmological Principle', in preparation.

Beisbart, C. and Jung, T. (2006), `Privileged, typical or not even that? Our place in the 
world according to the Copernican and cosmological principles',  {\em Journal of General 
Philosophy of Science} {\bf 37}, pp. 225-256.

Butterfield, J. (2007), `Stochastic Einstein Locality Revisited', {\em British Journal for 
Philosophy of Science} {\bf 58}, pp. 805-867.

Butterfield, J. (2011), `Laws, causation and dynamics at different levels', {\em Interface 
Focus} (The Royal Society, London) {\bf 2}, pp. pp. 101-114.

Butterfield, J. (2012), `Under-determination in cosmology: an invitation', {\em The 
Aristotelian Society Supplementary Volume 2012}, {\bf 86}, pp. 1-18.

Callender, C. (2004), `Measures, explanation and the past: should `special' initial 
conditions be explained?',  {\em British Journal for the Philosophy of Science} {\bf 55}, 
pp. 195-217.

Callender, C. (2011), `Hot and heavy matters in the foundations of statistical mechanics',  
{\em Foundations of Physics} {\bf 41}, pp. 960-981.

Carr, B. (2007), {\em Universe or Multiverse?}, Cambridge University Press.

Carrera M. and Giulini, D. (2010), `Influence of global cosmological expansion on local 
dynamics and kinematics', {\em Reviews of Modern Physics} {\bf 82}, pp. 169-208.

Clarkson, C., Ellis, G., Larena, J. and Umeh, O. (2011), `Does the growth of structure 
affect our dynamical models of the universe?', arxiv: astro-ph:1109.2314

Clifton, T., Clarkson C. and Bull P. (2011), `The isotropic blackbody CMB as evidence for a
homogeneous universe', arxiv: 1111.3794

Dicke, R. and Peebles, J. (1979), `The Big Bang cosmology---enigmas and nostrums', in S. 
Hawking and W. Israel eds., {\em General Relativity: an Einstein Centenary Survey}, 
Cambridge University Press; pp. 504-517.

Earman, J. (1995), {\em Bangs, Crunches, Whimpers and Shrieks}, Oxford University Press.

Earman, J. and Mosterin, J (1999), `A critical look at inflationary cosmology', {\em 
Philosophy of Science} {\bf 66}, pp. 1-49.

Ehlers, J., Geren P., and Sachs R. (1968), `Isotropic solutions of the Einstein-Liouville 
equations', {\em Journal of Mathematical Physics} {\bf 9}, pp. 1344-1349.

Ellis, G. (1975), `Cosmology and verifiability', {\em Quarterly Journal of the Royal 
Astronomical Society}, {\bf 16}, pp. 245-264.

Ellis, G. (1991), `Major themes in the relation between philosophy and cosmology', {\em 
Memorie della Societa Astronomica Italiana} {\bf 62}, pp. 553-605.

Ellis, G. (1993), `The physics and geometry of the universe: changing viewpoints', {\em 
Quarterly Journal of the Royal
Astronomical Society} {\bf 34}, pp. 315-330.

Ellis, G. (1999), `Before the beginning: emerging questions and uncertainties', {\em 
Astrophysics and Space Science} {\bf 269-270}, pp. 693-720.

Ellis, G. (1999a), `83 years in general relativity and cosmology: progress and problems', 
{\em Classical and Quantum Gravity} {\bf 16}, pp. A37-A75.

Ellis, G. (2007), `Issues in Philosophy of Cosmology', in part B of
J. Butterfield and J. Earman eds, {\em Philosophy of Physics}, Elsevier, volume 2 of the 
North Holland series, {\em The Handbook of Philosophy of Science}, pp. 1183-1286; 
astro-ph/0602280.

Ellis, G. (2011), `Inhomogeneity effects in cosmology', {\em Classical and Quantum Gravity} 
{\bf 28}, 164001.

Ellis, G. (2011a), `Does the multiverse really exist?', {\em Scientific American} {\bf 305}, 
August 2011, pp. 38-43.

Ellis, G. (2011b), `On the limits of quantum theory: contextuality and the quantum-classical 
cut', arxiv: quant-ph:1108.5261

Ellis, G., S. Nel, R. Maartens, W. Stoeger and A. Whitman (1985), `Ideal observational 
cosmology', {\em Physics Reports} {\bf 124}, pp. 315-417.

Glymour, D. (1977), `Indistinguishable spacetimes and the fundamental group', in  J. Earman, 
C. Glymour and J. Stachel (ed.s), {\em Foundations of Spacetime Theories}, Minnesota Studies 
in Philosophy of Science volume 8, University of Minnesota Press, pp. 50-60.

Guth, A. (1981), `Inflationary universe: a possible solution to the horizon and flatness 
problems', {\em Physical Review D} {\bf 23}, pp. 347-356.

Hawking, S. and Ellis, G. (1973), {\em The Large-scale Structure of Spacetime}, Cambridge 
University Press.

Hofer-Szabo, G., M. Redei and L. Szabo (2013), {\em The Principle of the Common Cause}, 
forthcoming, Cambridge University Press.

Kolb, E. and Turner, M. (1990), {\em The Early Universe}, Westview Press: Frontiers in 
Physics.

Kragh, H. (1996), {\em Cosmology and Controversy}, Princeton University Press.

Lahav, O. (2001), `Observational tests for the cosmological principle and world models', in 
R. Crittenden adn N. Turok (eds), {\em NATO ASIC Proc 565: Structure Formation in the 
Universe}, pp. 131f.

Lawrie, I. (1990), {\em A Unified Grand Tour of Theoretical Physics}, Institute of Physics 
Publishing, Adam Hilger.

Liddle, A. (2003), {\em An Introduction to Modern Cosmology}, John Wiley; second edition.

Linde, A. (1987), `Inflation and quantum cosmology', in S. Hawking and W. Israel eds., {\em 
300 Years of Gravitation}, Cambridge University Press; pp. 604-???.

Linde, A. (1990), {\em Particle Physics and Inflationary Cosmology}, Harwood Academic.

Linde, A. (2004), `Inflation, quantum cosmology and the anthropic principle', in J. Barrow, 
P. Davies and C. Harper eds., {\em Science and Ultimate Reality: quantum theory, cosmology 
and complexity}, Cambridge  University Press, pp. 426-458.

Longair, M. (2003), {\em Theoretical Concepts in Physics}, Cambridge University Press; 
second edition (first edition 1984).

Longair, M. (2006), {\em Cosmic Century}, Cambridge University Press.

MacCallum, M. (1979), `Anisotropic and inhomogeneous relativistic cosmologies', in S. 
Hawking and W. Israel eds., {\em General Relativity: an Einstein Centenary Survey}, 
Cambridge University Press; pp. 533-580.

Magnus, P. (2005), `Reckoning the shape of everything: underdetermination and 
cosmotopology', {\em British Journal for the Philosophy of Science} {\bf 56}, pp. 541-557.

Malament, D. (1977), `Observationally indistinguishable spacetimes', in  J. Earman, C. 
Glymour and J. Stachel (ed.s), {\em Foundations of Spacetime Theories}, Minnesota Studies in 
Philosophy of Science volume 8, University of Minnesota Press, pp. 61-80.

Manchak, J. (2009), `Can we know the global structure of spacetime?', {\em Studies in 
History and Philosophy of Modern Physics} {\bf 40}, pp. 53-56.

Manchak, J. (2011), `What is a physically reasonable spacetime?', {\em Philosophy of 
Science} {\bf 78}, pp. 410-420.

McMullin, E. (1985), `Galilean idealization',  {\em Studies in the History and Philosophy of 
Science} {\bf 16}, pp. 247-273.

McMullin, E. (1993), `Indifference principle and anthropic principle in cosmology', {\em 
Studies in the History and Philosophy of Science} {\bf 24}, pp. 359-389.

McMullin, E. (2005), `Anthropic explanation in cosmology', {\em Faith and Philosophy} {\bf 
22}, pp. 601-614.

McMullin, E. (2007), `Tuning fine-tuning', in J. Barrow et al. eds., {\em Fitness of the 
Cosmos for Life: Biochemistry and fine-tuning}, Cambridge University Press; pp. 70-94.

Misner, C., Thorne K. and Wheeler, J (1973), {\em Gravitation}, W.H. Freeman.

Nadathur, S., Hotchkiss, S. and Sarkar, S. (2011), `The integrated Sachs-Wolfe imprints of 
cosmic superstructures: a problem for $\Lambda$CDM', http://arxiv.org/abs/1109.4126v1.

Norton, J. (2011), `Observationally indistinguishable spacetimes: a challenge for any
inductivist', In G. Morgan, ed., {\em Philosophy of Science Matters: the
Philosophy of Peter Achinstein} Oxford University Press, 2011, pp. 164-176. Available at:\\
http://www.pitt.edu/~jdnorton/papers/Obs$\_$Equiv$\_$final.pdf

Penrose, R. (2004), {\em The Road to Reality}, Jonathan Cape.

Price, H. (2004), `On the origins of the arrow of time: why there is still a puzzle about 
the low entropy past', in C. Hitchcock ed. {\em Contemporary Debates in Philosophy of 
Science}, Oxford; Blackwell, pp. 219-239.

Reichenbach. H. (1956), {\em The Direction of Time}, University of California Press.

Rees, M. (1997), {\em Before the Beginning}, Simon and Schuster.

Rees, M. (2003),`Our complex cosmos and its future', in G. Gibbons, E. Shellard and S. 
Rankin (ed.s), {\em The Future of Theoretical Physics and Cosmology}, Cambridge University 
Press, pp. 17-37.

Rindler, W. (1977), {\em Essential Relativity}, Springer: second edition.

Rowan-Robinson, M. (1999), {\em The Nine Numbers of the Cosmos}, Oxford University Press.

Rowan-Robinson, M. (2004), {\em Cosmology}, Oxford University Press: fourth edition.

Rugh, S. and Zinkernagel, H. (2009), `On the physical basis of cosmic time', {\em Studies in 
History and Philosophy of Modern Physics} {\bf 40}, 1-19.

Rugh, S. and Zinkernagel, H. (2011), `Weyl's principle, cosmic time and quantum 
fundamentalism', in D. Dieks et al. (eds.), {\em Explanation, Prediction and Confirmation: 
the philosophy of science in a European perspective}, Springer, pp. 411-424.

Sarkar, S. (2007), `Is the evidence for dark energy secure?, 
http://arxiv.org/abs/0710.5307v2

Sarkar, P. et al (2009), `The scale of homogeneity of the galaxy distribution in SDSS DR6', 
{\em Monthly Notices of the Royal Astronomical Society} {\bf 399}, pp. L128-L131.

Saslaw, W. (1985), {\em Gravitational Physics of Stellar and Galactic Systems}, Cambridge 
University Press.

Saslaw, W. (2000), {\em The Distribution of the Galaxies}, Cambridge
University Press.

Sciama, D. (1971), {\em Modern Cosmology}, Cambridge University Press.

Silk, J. (1989),  {\em The Big Bang}, W.H. Freeman; revised and updated edition.

Silk, J. (2006), {\em The Infinite Cosmos}, Oxford University Press.

Smeenk, C. (2012), `Philosophy of cosmology', in {\em The Oxford Handbook of Philosophy of 
Physics}, ed. R. Batterman, Oxford University Press: pp. 607-652.

Smoot, G. (1993), {\em Wrinkles in Time: the imprint of creation}; (written with Keay 
Davidson), Little Brown and Company.

Steinhardt, P. (2011), `The inflation debate', {\em Scientific American} {\bf 304}, April 
2011, pp. 38-43.

Stoeger, W., Maartens, R., and Ellis, G. (1995), `Proving almost-homogeneity of the 
universe: an almost Ehlers, Geren and Sachs theorem', {\em Astrophysical Journal} {\bf 443}, 
pp. 1-5.

Torretti, R. (1983), {\em Relativity and Geometry}, Pergamon Press: reprinted by Dover.

Wald, R. (1984), {\em General Relativity}, University of Chicago Press.

Wallace, D. (2010), `Gravity, entropy and cosmology: in search of clarity',  {\em British 
Journal for the Philosophy of Science} {\bf 61}, pp. 513-540.

Wallace, D. (2011), `The logic of the past hypothesis',  forthcoming; available at: 
http://philsci-archive.pitt.edu/8894/

Weinberg, S. (1972), {\em Gravitation and Cosmology}, New York: John Wiley.

Weinberg, S. (2008), {\em Cosmology}, Oxford University Press.

Wilzcek, F. (2007), `Enlightenment, knowledge, ignorance, temptation', in B.Carr (ed.), {\em 
Universe or Multiverse?}, Cambridge University Press, pp. 43-54. arXiv: hep-ph/0512187v2

Yang, A. and Saslaw, W. (2011), `The galaxy counts-in-cells distribution from the SDSS', 
{\em Astrophysical Journal} {\bf 729}, 123: arXiv:1009.0013v2. 
doi:10.1088/0004-637X/729/2/123

Zinkernagel, H. (2002), `Cosmology, particles and the unity of science', {\em Studies in 
History and Philosophy of Modern Physics} {\bf 33B}, pp. 493-516.

Zinkernagel, H. (2011), `Some trends in the philosophy of physics', {\em Theoria} (Spain) 
{\bf 26}, pp. 215-241; available at: http://philsci-archive.pitt.edu/8761/

\end{document}